\documentclass[journal=NanoLetters,manuscript=article]{achemso}

\setkeys{acs}{keywords = true}
\usepackage[version=3]{mhchem} 
\usepackage[T1]{fontenc}       
\usepackage{multirow}

\author{Xue Zhou}
\affiliation{State Key Laboratory of Low Dimensional Quantum Physics and Department of Physics, Tsinghua University Beijing 100084, People$\prime$s Republic of China}
\author{Zeyu Jiang}
\affiliation{State Key Laboratory of Low Dimensional Quantum Physics and Department of Physics, Tsinghua University Beijing 100084, People$\prime$s Republic of China}
\author{Kenan Zhang}
\affiliation{State Key Laboratory of Low Dimensional Quantum Physics and Department of Physics, Tsinghua University Beijing 100084, People$\prime$s Republic of China}
\author{Wei Yao}
\affiliation{State Key Laboratory of Low Dimensional Quantum Physics and Department of Physics, Tsinghua University Beijing 100084, People$\prime$s Republic of China}
\author{Mingzhe Yan}
\affiliation{State Key Laboratory of Low Dimensional Quantum Physics and Department of Physics, Tsinghua University Beijing 100084, People$\prime$s Republic of China}
\author{Hongyun Zhang}
\affiliation{State Key Laboratory of Low Dimensional Quantum Physics and Department of Physics, Tsinghua University Beijing 100084, People$\prime$s Republic of China}
\author{Wenhui Duan}
\affiliation{State Key Laboratory of Low Dimensional Quantum Physics and Department of Physics, Tsinghua University Beijing 100084, People$\prime$s Republic of China}
\author{Shuyun Zhou}
\affiliation{State Key Laboratory of Low Dimensional Quantum Physics and Department of Physics, Tsinghua University Beijing 100084, People$\prime$s Republic of China}
\altaffiliation{Collaborative Innovation Center of Quantum Matter, Beijing, People$\prime$s Republic of China}
\email{syzhou@mail.tsinghua.edu.cn}

\title[An \textsf{achemso} demo]
{Electronic structure of molecular beam epitaxy grown 1T$^\prime$-MoTe$_2$ film and strain effect}

\begin{document}


\begin{abstract}
Atomically thin transition metal dichalcogenide films with distorted trigonal (1T$^\prime$) phase have been predicted to be candidates for realizing quantum spin Hall effect. Growth of 1T$^\prime$ film and experimental investigation of its electronic structure are critical. Here we report the electronic structure of 1T$^\prime$-MoTe$_2$ films grown by molecular beam epitaxy (MBE).  Growth of the 1T$^\prime$-MoTe$_2$ film depends critically on the substrate temperature, and successful growth of the film is indicated by streaky stripes in the reflection high energy electron diffraction and sharp diffraction spots in low energy electron diffraction.  Angle-resolved photoemission spectroscopy measurements reveal a metallic behavior in the as-grown film with an overlap between the conduction and valence bands. First principles calculation suggests that a suitable tensile strain along the a-axis direction is needed to induce a gap to make it an insulator. Our work not only reports the electronic structure of MBE grown 1T$^\prime$-MoTe$_2$ films, but also provides insights for strain engineering to make it possible for quantum spin Hall effect.

\end{abstract}


\section{Introduction}
Topological materials have provided an important platform for exploring new physics and realizing novel quantum phenomena.\cite{Hasan2010,ZhangSC2011} For example,  quantum spin Hall effect (QSHE)\cite{ZhangSC2006,ZhangSC2007,ZhangSC2008} is expected in two-dimensional topological insulators.\cite{Mele2005} Transition metal dichalcogenides (TMDCs) with distorted trigonal structure (1T$^\prime$) have been predicted to be important candidates for realizing QSHE with potential applications in topological field effect transistors.\cite{FuL2014, FengSP2012, AbidAA2017, LuoWD2018, YuGL2017} Recently, 1T$^\prime$-WTe$_2$ thin films have been revealed to show electronic properties compatible with QSHE.\cite{ShenZX2017,XX2017,Herrero2018} 1T$^\prime$-MoTe$_2$ has similar crystal structure to 1T$^\prime$-WTe$_2$\cite{Lee2015}  and can also be a potential candidate for QSHE. MoTe$_2$ crystalizes in three structures, hexagonal (2H), monoclinic (1T$^\prime$)\cite{Lee2015} and orthorhombic (T$_d$)\cite{Zhou2016}. Bulk single crystal of 1T$^\prime$-MoTe$_2$ undergoes a phase transition to T$_d$ phase\cite{Zhou2016} which hosts type-II Weyl fermions\cite{Bernevig2015, Zhousy2016,Adam2016} and a superconducting transition has been reported at even lower temperature,\cite{Medvedev2016} however, mechanically exfoliated few layered 1T$^\prime$-MoTe$_2$ has been reported to be a semiconductor.\cite{Lee2015}

While growth of 1T$^\prime$-MoTe$_2$ thin film has been reported by chemical vapor deposition,\cite{LeeYH2015,KongJ2016,Killawa2016,LeeYH2016,LiuZ2016} molecular beam epitaxy (MBE) growth of 1T$^\prime$-MoTe$_2$ films under ultra-high vacuum has the advantage of being directly compatible with {\it in situ} electronic structure measurement by angle-resolved photoemission spectroscopy (ARPES). So far, despite extensive efforts, molecular beam epitaxy  growth of atomically thin 1T$^\prime$-MoTe$_2$ films\cite{XieM2017,ZhangXA2017,XingHG2017,ShenZX2018} has been challenging due to the existence of another stable phase 2H-MoTe$_2$,\cite{Batzill2015} which often leads to a mixture of both 1T$^\prime$ and 2H phases in as-grown MoTe$_2$ films.\cite{XieM2017,ZhangXA2017,XingHG2017} Here we report the successful growth of 1T$^\prime$-MoTe$_2$ films at the optimum growth condition after a systematic study of the film growth at different substrate temperatures.  ARPES measurements show that the as-grown film shows a metallic behavior with an overlap between the conduction and valence bands. First principles calculation suggests that a 3\% tensile uniaxial strain along the a-axis direction is needed to induce a significant gap to be compatible with QSHE.

\section{Methods}
1T$^\prime$-MoTe$_2$ films were grown on bilayer graphene/6H-SiC (0001) by MBE. The 6H-SiC(0001) substrate was degassed at 650 $^\circ \mathrm{C}$ and annealed from 650 to 1350 $^\circ \mathrm{C}$ for 60 cycles to form bilayer graphene films on the top surface.\cite{XueQK2013} High purity Mo (99.99 \% purity) and Te (99.999 \% purity) were then evaporated through an e-beam evaporator and Knudsen cell respectively with a flux ratio of $\sim$ 1:20. The growth process is monitored by \emph{in situ} reflection high energy electron diffraction (RHEED) and low energy electron diffraction (LEED). ARPES measurements were performed \emph{in situ} with a Helium lamp source at a temperature of $\sim$10 K under ultra-high vacuum.

The density functional theory (DFT) calculations are performed using the Vienna ab initio simulation package (VASP)\cite{Fur1996} with the Perdew-Burke-Ernzerhof (PBE)\cite{Mat1996}  exchange-correlation functional and a plane wave energy cut-off of 500 eV. A k-point grid of 16 $\times$ 20 $\times$ 1 is applied to sample the Brillouin zone. The pristine geometric structure of monolayer is fully relaxed until the residual forces on each atom are less than 0.001 eV/\r{A}, and the obtained equilibrium lattice parameters are a = 3.475 \r{A} and b = 6.367 \r{A}. To simulate the uniaxial strain along the a-axis (b-axis), a stain is applied along the a-axis (b-axis), and the length of b-axis (a-axis) as well as the ionic positions is optimized until the residual forces are less than 0.001 eV/\r{A}. The spin-orbit coupling (SOC) effect has been taken into account in our calculations.

\section{Results and discussion}

\begin{center}
	
	\includegraphics[width=16.8 cm] {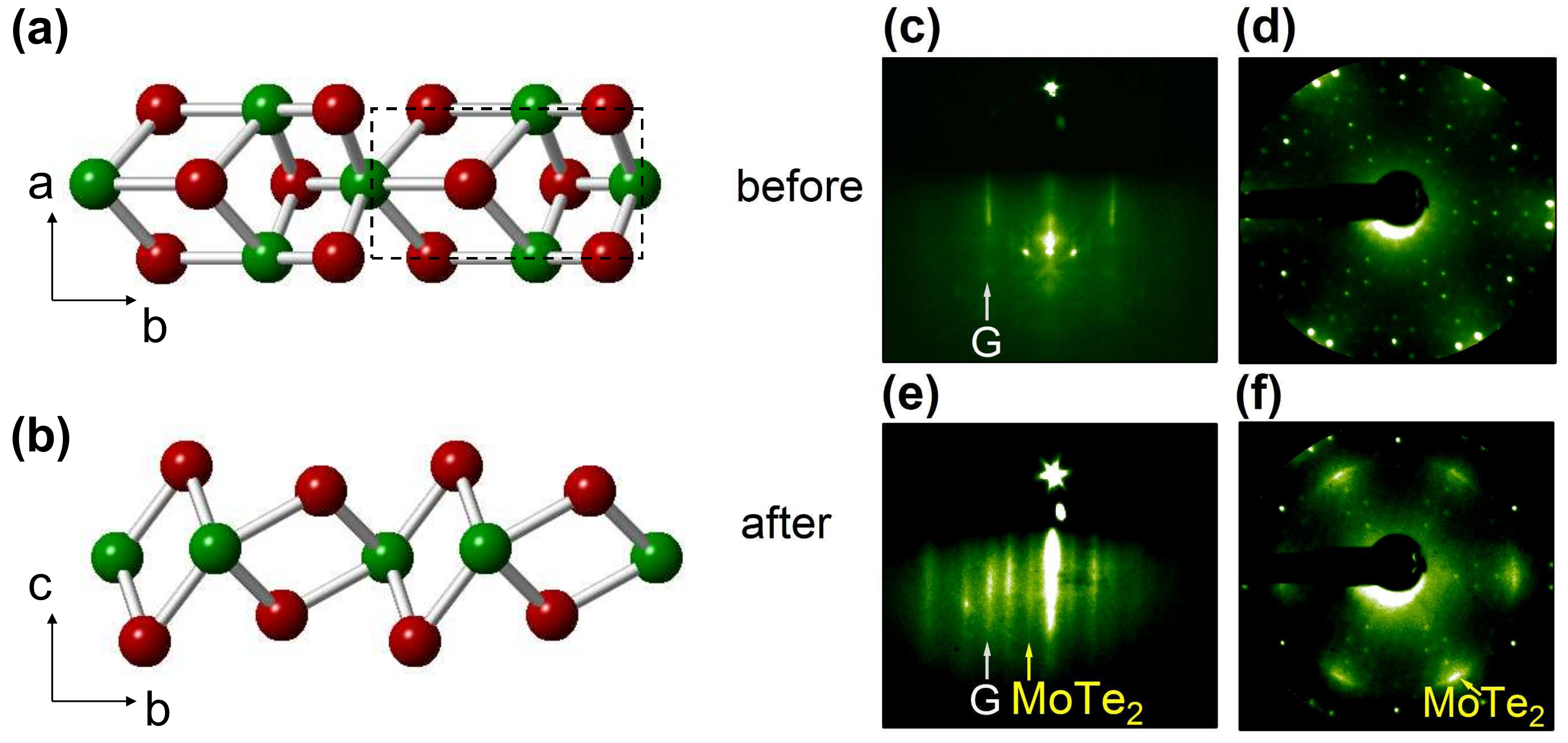}\\[5pt]
	\parbox[c]{15.0cm}\footnotesize{\bf Fig.~1.} {Crystal structure of 1T$^\prime$-MoTe$_2$ and diffraction patterns from RHEED and LEED. (a)-(b) Top and side views of the crystal structure of 1T$^\prime$-MoTe$_2$. Red and green balls represent Te and Mo atoms, respectively.  The dashed box indicates the unit cell. (c), (e) RHEED patterns of the graphene/SiC substrates (c) and the as-grown 1T$^\prime$-MoTe$_2$ sample (e). (d), (f) LEED patterns of graphene/SiC substrates (d) and the as-grown 1T$^\prime$-MoTe$_2$ sample (f) measured at a beam energy of 120 eV. The white and yellow arrows indicate the patterns from graphene and 1T$^\prime$-MoTe$_2$, respectively.}
	
\end{center}
Figure 1(a) and 1(b) show the top and side views of the crystal structure of 1T$^\prime$-MoTe$_2$. The Mo atoms deviate from the center of the octahedra formed by six Te atoms, forming zigzag Mo chains along the a-axis direction (see the top view in Fig. 1(a)) and distorted Te octahedra in the b-c plane (side view in Fig. 1(b)).  Graphene is a fantastic substrate for growing films with different crystal structures and symmetries through van der Waals epitaxiy,\cite{KomaA1984} and is used as the substrate for growing 1T$^\prime$-MoTe$_2$ film.
Figure 1(c) and 1(d) show the RHEED and LEED of the graphene/SiC substrate. Figure 1(e) and 1(f) show the RHEED and LEED patterns of 1T$^\prime$-MoTe$_2$ films under optimum growth conditions. Sharp streaky stripes (indicated by yellow arrow in Fig. 1(e)) and six diffraction spots (Fig. 1(f)) from the 1T$^\prime$-MoTe$_2$ film are observed in the RHEED and LEED patterns respectively. The 1T$^\prime$-MoTe$_2$ film grows mainly along the same orientation as the graphene substrate with a small distribution of azimuthal angles in the LEED pattern due to the weak van der Waals growth with weak coupling between 1T$^\prime$-MoTe$_2$ film and graphene. Because of the different crystal symmetries between the substrate (three fold symmetry) and the 1T$^\prime$-MoTe$_2$ film (two fold symmetry), there are three equivalent orientations of 1T$^\prime$-MoTe$_2$ films on graphene, leading to apparently hexagonal LEED patterns, similar to the case of 1T$^\prime$-WTe$_2$ film\cite{ShenZX2017} and previous report on 1T$^\prime$-MoTe$_2$\cite{ShenZX2018} yet with better LEED pattern. The observation of diffraction spots from both the 1T$^\prime$-MoTe$_2$ film and the graphene substrate suggests that the 1T$^\prime$-MoTe$_2$ film is atomically thin,  $\sim$ 1 ML thick. Increasing the growth time leads to weaker diffraction spots from the substrate and graphene diffraction spots disappear at 2ML (Supplemental Material ), however, no major change in the electronic structure is observed since the difference in the electronic structure of monolayer, bilayer and multilayer 1T$^\prime$-MoTe$_2$ films is small due to the small band splitting.  Using the lattice constants of graphene as a reference, the extracted in-plane lattice constants of 1T$^\prime$-MoTe$_2$ from the LEED pattern are a = 3.47 \r{A} and b = 6.48 \r{A}, suggesting a 2\% (tensile) strain along the b-axis direction compared to the lattice constants of a = 3.48 \r{A} and b = 6.33 \r{A} in the bulk crystal.\cite{Lee2015}

\begin{center}
	
	\includegraphics[width=16.8 cm] {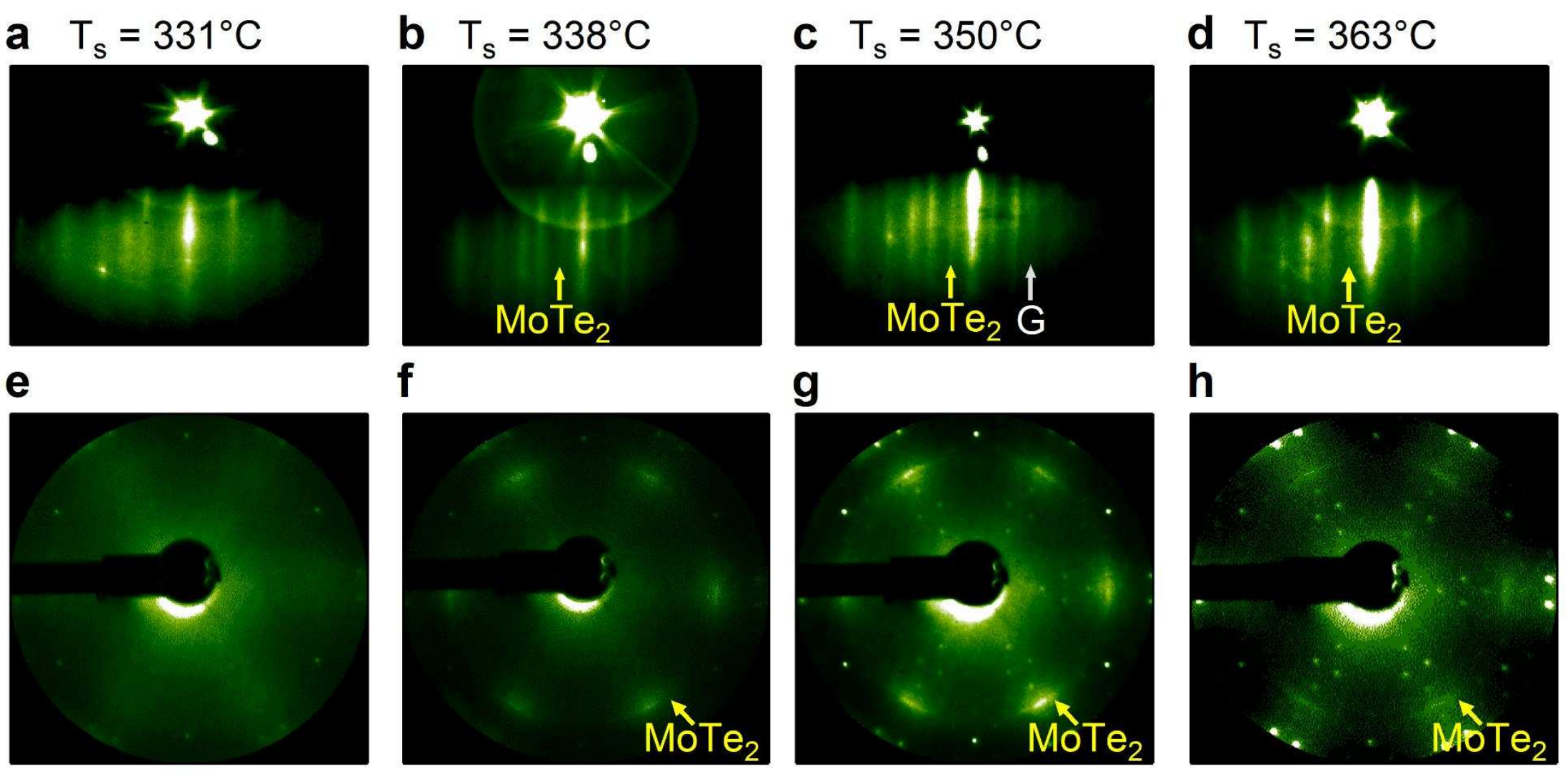}\\[5pt]
	\parbox[c]{15.0cm}\footnotesize{\bf Fig.~2.} {RHEED and LEED patterns for films grown at different substrate temperatures. (a)-(d) RHEED patterns after growing at the substrate temperature of 331$^\circ \mathrm{C}$, 338 $^\circ \mathrm{C}$, 350 $^\circ \mathrm{C}$ and 363 $^\circ \mathrm{C}$ respectively. (e)-(h) Corresponding LEED patterns. The yellow and white arrows indicate diffraction spots from 1T$^\prime$-MoTe$_2$ and graphene.}
	
\end{center}

The growth condition is critically dependent on the substrate temperature. Figure 2 shows a systematic study of the RHEED and LEED patterns of films grown at different substrate temperatures while maintaining other experimental conditions fixed. When the substrate temperature is 331 $^\circ \mathrm{C}$, no detectable signals from MoTe$_2$ are observed in the RHEED (Fig. 2(a)) or LEED patterns (Fig. 2(e)). Only in a small temperature window of $\sim$ 25 $^\circ \mathrm{C}$ between 338 $^\circ \mathrm{C}$ and 363 $^\circ \mathrm{C}$, streaky stripes can be observed (indicated by yellow arrows) in RHEED (Fig. 2(b)-(d)) and diffraction spots are observed in the LEED patterns (Fig. 2(f)-(h)). The sharpest streaky stripes from the RHEED pattern (Fig. 2(c)) and the best signal from the LEED pattern (Fig. 2(g)) at 350  $^\circ \mathrm{C}$ of the substrate temperature indicate that the optimum growth condition for the substrate temperature is 350  $^\circ \mathrm{C}$.

\begin{center}
	
	\includegraphics[width=16.8 cm] {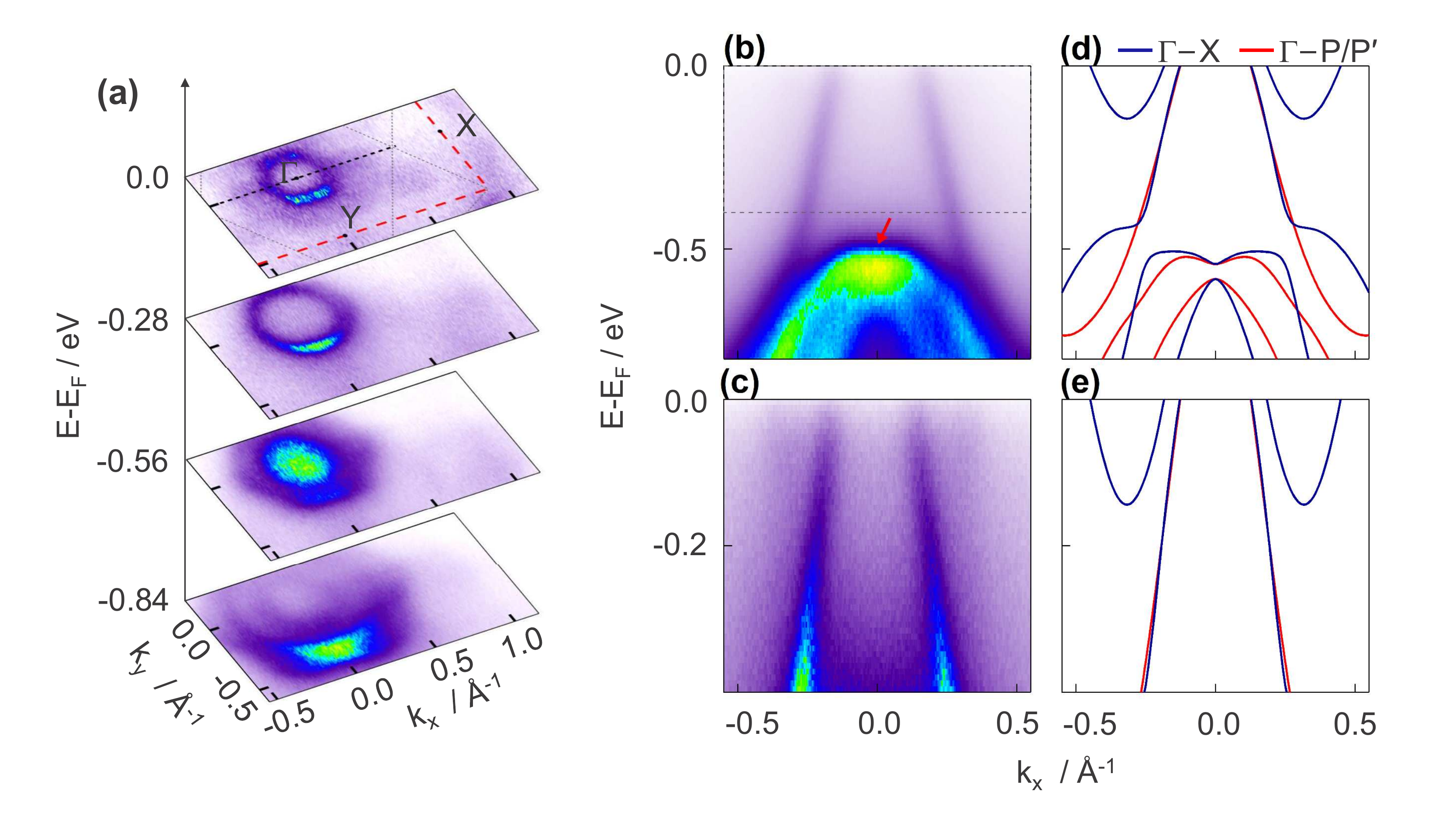}\\[5pt]
	\parbox[c]{15.0cm}\footnotesize{\bf Fig.~3.} {Electronic structure of 1T$^\prime$-MoTe$_2$ revealed by ARPES measured at $\sim$ 10 K. (a) Intensity maps measured from E$_F$ to -0.84 eV with an integrated energy window of 50 meV. (b) ARPES spectrum measured along the $\Gamma$-X direction (marked by dotted line in (a)). (c) Zoom-in dispersion near E$_F$. (d)-(e) Calculated dispersions along the $\Gamma$-X direction (blue) and $\Gamma$-P/P$^\prime$ (red) (d) and zoom-in dispersion near E$_F$ (e) with a uniaxial strain of 2\% along the b-axis direction and a shift -0.09 eV in energy.}
	
\end{center}

Although bulk crystal of 1T$^\prime$-MoTe$_2$ is quite stable in air, atomically thin 1T$^\prime$-MoTe$_2$ films have been reported to be very sensitive to air.\cite{Killawa2016,XieM2017,SchaakER2016,Batzill2015,HeinzTF2014} Therefore {\it ex situ} atomic force microscopy (AFM) and Raman characterizations of the films are not practical. Direct experimental electronic structure of the as-grown film by {\it in situ} ARPES measurements can provide direct evidence for the 1T$^\prime$-MoTe$_2$ films, and moreover, to provide insights for evaluating its compatability with QSHE.

Figure 3 shows the electronic structure of 1T$^\prime$-MoTe$_2$ revealed by \emph{in situ} ARPES measurements. Figure 3(a) shows intensity maps measured from E$_F$ to -0.84 eV. Since the 2H-MoTe$_2$ is a semiconductor with a large gap of more than 1 eV,\cite{Zhousy2018} this confirms that the measured band dispersion is not from the 2H-MoTe$_2$ but from the 1T$^\prime$-MoTe$_2$. A hexagonal pocket centered at the $\Gamma$ point is observed at E$_F$ and its size increases at low energies, suggesting that it is a hole pocket. Below -0.56 eV, a new pocket emerges at the $\Gamma$ point and further splits into two circular pockets, resulting in three hole pockets in total at -0.84 eV.  Figure 3(b) shows the dispersions measured along the $\Gamma$-X direction. The dispersion shows a linear dispersing hole pocket through E$_F$, and two parabolic bands below -0.56 eV, consistent with the intensity maps in Fig. 3(a). Figure 3(c) shows a zoom-in of the dispersion near E${_F}$. In addition to the hole pocket near E${_F}$ as discussed above, there is another dispersing band within -0.15 eV, suggesting that this is likely the electron pocket from the conduction band. Figure 3(d) and 3(e) show the calculated band dispersion for comparison.\cite{ShenZX2017}  By using the extracted experimental lattice constant from LEED, which indicates a strain of 2\% (tensile) along the b-axis direction, a good agreement with the experimental results is obtained. The overall band structure is similar to that of 1T$^\prime$-WTe$_2$,\cite{ShenZX2017,FengJ2016} yet with two major differences. Firstly, the two bands below -0.56 eV almost cross (pointed by red arrow in Fig. 3(b)), which is different from those in 1T$^\prime$-WTe$_2$ film and is not discussed in previous work in 1T$^\prime$-MoTe$_2$.\cite{ShenZX2017} Indeed, the better agreement between the ARPES data and the calculated band structure of strained film compared to calculated result of unstrained film in previous work\cite{ShenZX2017} also indicates the important role of stain in this material. Secondly, both the experimental and calculated results reveal an overlap between the conduction and valence bands, while doped 1T$^\prime$-WTe$_2$ has been reported to be an insulator with a gap of 45 meV.\cite{ShenZX2017} Therefore, our ARPES data and calculation show that different from 1T$^\prime$-WTe$_2$, the as-grown 1T$^\prime$-MoTe$_2$ film is metallic with an overlap between the valence and conduction bands.

\begin{center}
	
	\includegraphics[width=16.8 cm] {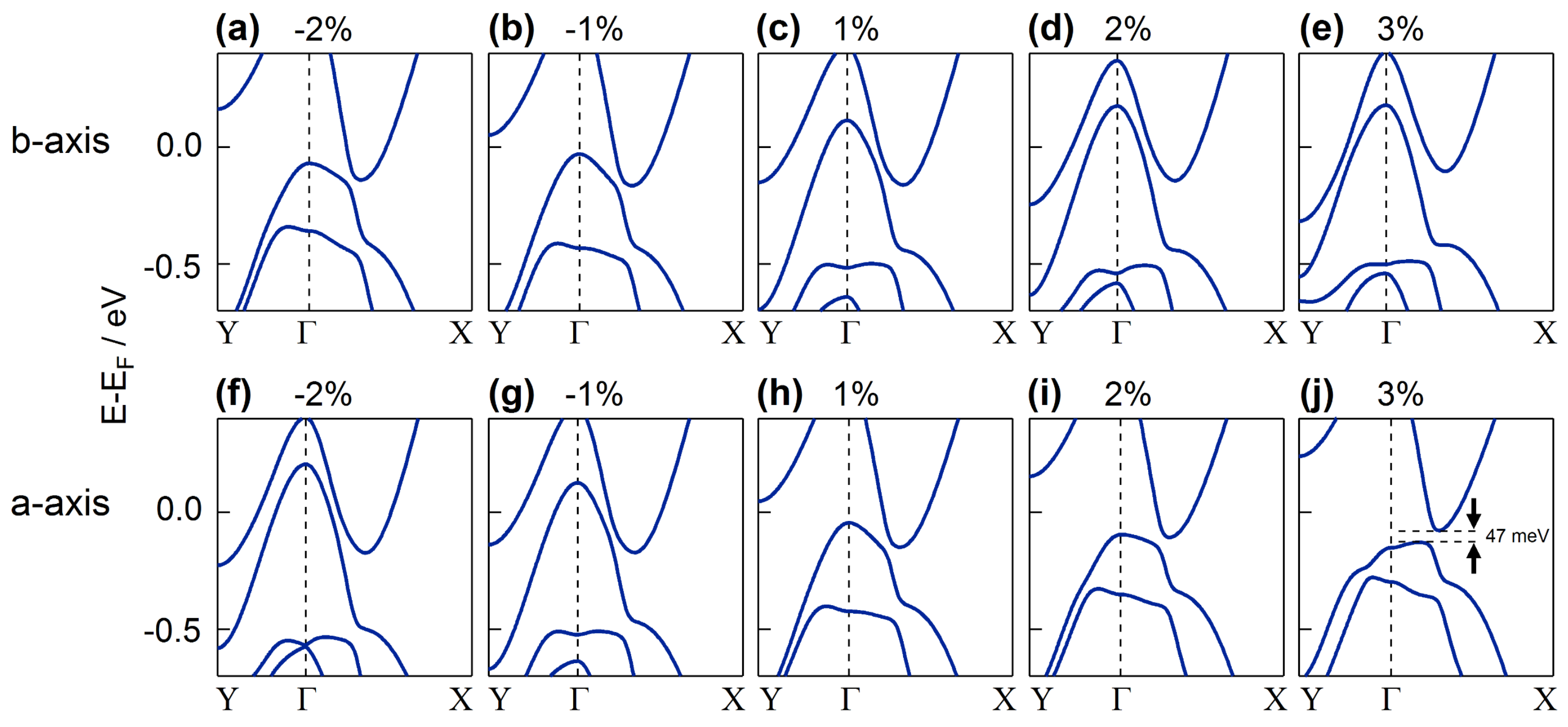}
	\parbox[c]{15.0cm}\footnotesize{\bf Fig.~4.} {Evolution of the electronic structure with strain from first principles calculation. (a)-(e) Calculated band structure of monolayer 1T$^\prime$-MoTe$_2$ with uniaxial strains from -2\% to 3\% along the b-axis direction. (f)-(j) Calculated band structure of monolayer 1T$^\prime$-MoTe$_2$ with uniaxial strains from -2\% to 3\% along the a-axis direction. All calculated dispersions are shifted by -0.09 eV in energy.}
	
\end{center}

Since the electronic structures of 1T$^\prime$-MoTe$_2$ films\cite{FuL2014} and bulk crystals\cite{Despina2017, Woo2017, QinSS2016} are strongly dependent on the strain, we calculate the evolution of the electronic structure with uniaxial strain to provide more insights. Figure 4 shows the calculated band structure of monolayer 1T$^\prime$-MoTe$_2$ film under uniaxial strain along the b-axis (Fig. 4(a)-(e)) and a-axis (Fig. 4(f)-(j)) with strains ranging from -2\% to 3\%. The application of a tensile strain along the a-axis direction has similar effect to the application of a compressive strain along the b-axis direction. The uniaxial strain has two major effects. Firstly, it changes the splitting of the two bands below -0.5 eV at the $\Gamma$ point. More importantly,  it changes the energy position of these valence bands significantly, while maintaining the energy position of the conduction band. By applying a tensile strain along the a-axis, the overlap between the valence and conduction bands decreases, until eventually a gap of 47 meV emerges at 3\% strain (Fig. 4(j)). The opening of such a band gap is critical, since it makes it potentially a quantum spin Hall insulator if the Fermi energy is further tuned to inside the gap region. Therefore in order to realize QSHE in 1T$^\prime$-MoTe$_2$ films, a tensile strain (3\%) along the a-axis is needed.

\section{Conclusion}
To summarize, we have successfully grown high-quality atomically thin 1T$^\prime$-MoTe$_2$ films using MBE after a systematic investigation of the growth at different substrate temperatures, which is confirmed by RHEED, LEED and ARPES measurements. Furthermore, ARPES measurements show that the as-grown film is a metal with an overlap between the conduction and valence bands, which is attributed to strain effect. Comparison of calculated band structure at different strains further suggests that a suitable tensile strain (3\% tensile strain along the a-axis direction) can induce a significant gap between the conduction and valence bands. Our work not only reports the MBE growth conditions for obtaining  1T$^\prime$-MoTe$_2$ thin film and its experimental electronic structure, but also provides insights for band structure engineering of 1T$^\prime$-MoTe$_2$ film to make it a quantum spin Hall insulator.

\addcontentsline{toc}{chapter}{Appendix A: Supplemental Material}
\section*{Appendix A: Supplemental Material}
Figure A1 shows a comparison of RHEED, LEED and calculated electronic structure for 1 ML and 2 ML film. The observation of graphene spots in Fig. A1(b) shows that the sample is $\sim$ 1ML. By doubling the growth time, the diffraction spots from graphene disappears. A comparison of the electronic structure for 1 ML and 2 ML shows that the splitting is very weak and beyond the resolution of ARPES experiments.

\begin{center}
	
	\includegraphics[width=16.8 cm] {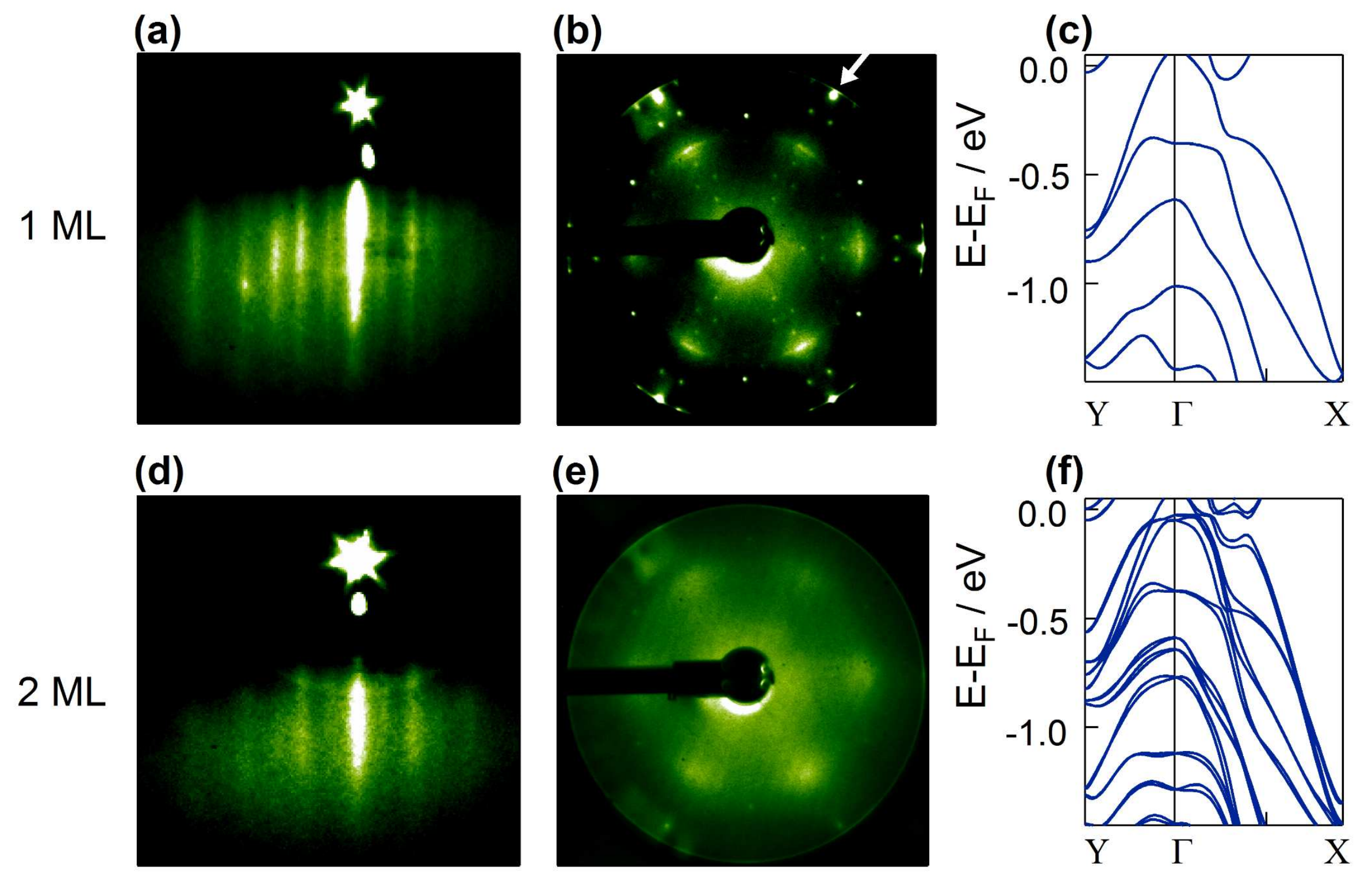}\\[5pt]
	\parbox[c]{15.0cm}\footnotesize{\bf Fig.~A1.} {comparison of RHEED, LEED and calculated electronic structure for 1 ML and 2 ML film. (a)-(c) RHEED, LEED and calculated electronic structure of 1 ML film, white arrow in (b) indicates the patterns from graphene. (d)-(f) RHEED, LEED and calculated electronic structure of 2 ML film.}
	
\end{center}


\begin{acknowledgement}
	This work is supported by the Ministry of Science and Technology of China (Grant No.  2016YFA0301004 and 2015CB921001) and the National Natural Science Foundation of China (Grant No. 11334006, 11725418, 11674188).
\end{acknowledgement}



\providecommand{\newblock}{}

\end{document}